\documentclass[showpacs,aps,prb,twocolumn,superscriptaddress,longbibliography]{revtex4-1}

\usepackage{times}
\usepackage{mathptmx}
\usepackage{graphicx}
\usepackage{subfigure}
\usepackage{bm}
\usepackage{color}

\usepackage{amsmath}
\usepackage{amssymb}
\usepackage{enumerate}
\usepackage{xspace}
\usepackage{mathrsfs}

\newcommand{\smJij}{\ensuremath{J_{ij}}\xspace}
\newcommand{\smmu}{\ensuremath{\mu_{\mathrm{s}}}\xspace}
\newcommand{\smH}{\ensuremath{\mathbf{H}}\xspace}

\newcommand{\sms}{\ensuremath{\mathbf{S}}\xspace}
\newcommand{\HH}{\ensuremath{\mathscr{H}}}
\newcommand{\smku}{\ensuremath{k_{\mathrm{u}}}}
\newcommand{\vampire}{\textsc{vampire} }
\newcommand{\Tc}{\ensuremath{T_{\mathrm{c}}}\xspace}

\newcommand{\kB}{\ensuremath{k_{\mathrm{B}}}\xspace}
\newcommand{\muB}{\ensuremath{\mu_{\mathrm{B}}}\xspace}
\newcommand{\Tsim}{\ensuremath{T_{\mathrm{sim}}}\xspace}
\newcommand{\Texp}{\ensuremath{T_{\mathrm{exp}}}\xspace}

\usepackage{xcolor}
\usepackage{soul}
\usepackage{soulpos}

\begin{document}


\title{Quantitative simulation of temperature dependent magnetization\\ dynamics and equilibrium properties of elemental ferromagnets}

\author{R. F. L. Evans}
\email{richard.evans@york.ac.uk}
\affiliation{Department of Physics, The University of York, York, YO10 5DD, UK}
\author{U. Atxitia}
\affiliation{Department of Physics, The University of York, York, YO10 5DD, UK}
\affiliation{Fachbereich Physik and Zukunftskolleg, Universit\"{a}t Konstanz, D-78457 Konstanz, Germany}
\author{R. W. Chantrell}
\affiliation{Department of Physics, The University of York, York, YO10 5DD, UK}
\begin{abstract}
Atomistic spin model simulations are immensely useful in determining temperature dependent magnetic properties, but are known to give the incorrect dependence  of the magnetization on temperature compared to experiment owing to their classical origin. We find a single parameter rescaling of thermal fluctuations which gives quantitative agreement of the temperature dependent magnetization between atomistic simulations and experiment for the elemental ferromagnets Ni, Fe , Co and Gd. 
Simulating the sub-picosecond magnetization dynamics of Ni under the action of a laser pulse we also find quantitative agreement with experiment in the ultrafast regime. This enables the quantitative determination of temperature dependent magnetic properties allowing for accurate simulations of magnetic materials at all temperatures.
\end{abstract}

\pacs{75.30.Kz,75.78.-n,75.10.Hk,75.30.Ds}
\maketitle



\section{Introduction}
Magnetic materials are used in a wide range of technologies with applications in power generation\cite{GutfleischAM2011}, data storage\cite{Plumer2001,ParkinScience2008}, data processing\cite{WolfScience2001}, and cancer therapy\cite{JordanJMMM1999}. All of these magnetic technologies operate at a wide range of temperatures, where microscopic thermal fluctuations determine the thermodynamics of the macroscopic magnetic properties. Recently thermal fluctuations in the magnetization have been shown to drive not only a number of phenomena  of great fundamental interest, for example ultrafast demagnetization\cite{BeaurepairePRL1996}, thermally induced magnetic switching\cite{RaduNature2011,OstlerNatCom2012}, spin caloritronics\cite{BauerNMat2012}  but also next generation technologies such as heat assisted magnetic recording\cite{KryderIEEE2008} and thermally assisted magnetic random access memory\cite{PrejbeanuJPCM2007}. Design requirements for magnetic devices typically require complex combinations of sample geometry, tuned material properties and dynamic behavior to optimize their performance. Understanding the complex interaction of these physical effects often requires numerical simulations such as those provided by micromagnetics\cite{ScholzMAGPAR2003,FischbacherNMAG2007,ChangFASTMAG2011} or atomistic spin models\cite{EvansVMPR2013}. Micromagnetic simulations at elevated temperatures\cite{GaraninPRB1997,EvansPRB2012} in addition need the temperature dependence of the main parameters\cite{KazantsevaPRB2008} such as the magnetization, micromagnetic exchange\cite{AtxitiaEXC2010} and effective anisotropy\cite{AsselinCMC2010}. Although analytical approximations for these parameters exist, multiscale \textit{ab-initio}/atomistic simulations \cite{MyrasovFePtEPL2005,KazantsevaPRB2008} have been shown to more accurately determine them.

With atomistic simulations the disparity between the simulated and  experimental temperature dependent magnetization curves arises due to the classical nature of the atomistic spin model\cite{KuzminPRL2005}. At the macroscopic level the temperature dependent magnetization is well fitted by the phenomenological equation proposed by Kuz'min\cite{KuzminPRL2005}. However, the Kuz'min equation merely describes the form of the curve with little relation to the microscopic interactions within the material which determine fundamental properties such as the Curie temperature. Ideally one would perform \textit{ab-initio} 3D quantum Monte Carlo simulations\cite{SandvikPRB1991}. Although this is possible for a small number of atoms, for larger ensembles the multiscale approach using atomistic models parameterized with \textit{ab-initio} information remains the only feasible way to connect the quantum and thermodynamic worlds. At the same time there is a pressing need to match parameters determined from the multiscale model to experiment to understand complex temperature dependent phenomena and magnetization dynamics. Atomistic models also provide a natural way to model non-equilibrium temperature effects such as ultrafast laser-induced magnetization dynamics\cite{BeaurepairePRL1996,RaduNature2011,OstlerNatCom2012} or quasi-equilibrium properties such as the Spin-Seebeck effect created by temperature gradients\cite{HinzkePRL2011,BauerNMat2012}. Alternative numerical\cite{KormannPRB2010,KormannPRB2011} and analytical\cite{Halilov1997,Halilov1998} approaches have been used to successfully describe the low temperature behavior, but add significant complexity compared to simple classical simulations.

In this work we present a single parameter rescaling of thermal fluctuations within the classical Heisenberg model which correctly describes the equilibrium magnetization at all temperatures. Since the temperature dependence of important magnetic properties such as anisotropy and exchange often arises due to fluctuations of the magnetization, this rescaling can also be used to accurately calculate their temperature variation. Furthermore we show that this rescaling is capable of quantitatively describing ultrafast magnetization dynamics in Ni. The quantitative agreement of the magnetic properties between theory and experiment enables the next generation of computer models of magnetic materials accurate for all temperatures and marks a fundamental step forward in magnetic materials design.

\section{Form of the temperature dependent magnetization}
We first consider the physics behind the form of $M(T)$. Atomistic spin dynamics (ASD) considers localized classical atomic spins $\mathbf{S}_i=\mu_s \mathbf{s}_i$ where $\mu_s$ is the magnetic moment, \textit{i.e} the spin operator $\mathbf{S}_i$ at each lattice site takes unrestricted values on the unit sphere surface $|\mathbf{s}_i|=1$  whereas in the quantum case they are restricted to their particular eigenvalues. However, when calculating the macroscopic thermodynamic properties of a many spin system, as ASD eventually does, this distinction is not apparent since the mean value of $\langle \mathbf{S} \rangle=M(T)$ is not restricted to quantized values within the quantum description.

A direct consequence of the distinction between classical and quantum models is manifest in the particular statistical properties of each approach. As is well-known, thermal excitation of the spin waves in ferromagnets leads to a decrease of the macroscopic magnetization $M(T)$ as temperature increases.\cite{Bloch1930} In the limit of low temperatures, $m(T)=M(T)/M(0)$ can be calculated as $m=1-\rho(T)$, where 
$\rho(T)=(1/\mathcal{N}) \sum_{\bm{k}} n_{\bm{k}}$ is the sum over the wave vector $\bm{k}$ of the spin wave occupation number in  the Brillouin zone\cite{BastardisPRB2012,QuantumTheoryMagnetism2007}.

The occupation number of a spin wave of energy $\epsilon_k$ corresponds to the high temperature limit of the Boltzmann law in reciprocal space,\cite{BastardisPRB2012} $n_{\bm{k}}=k_B T/\varepsilon_{\bm{k}}$, where $T$ is the temperature, $k_B$ is the Boltzmann constant, while quantum spin waves follow the Bose-Einstein distribution ($n_{\bm{k}}=1/\left(\exp(\varepsilon_{\bm{k}}/k_B T))-1\right)$).
Different forms of $m(T)$  are expected due to  the specific $n_{\bm{k}}$ used in each picture.

Given that the spin wave energies $\varepsilon_{\bm{k}}$ are the same in both the quantum and classical model the difference in the form of the $M(T)$ curve comes solely from the different statistics.
We can illustrate the difference in the statistics by considering the simplest possible ferromagnet described by a quantum and classical spin Heisenberg Hamiltonian. To do so, we consider the anisotropy and external magnetic fields as small contributions to the Hamiltonian in comparison to the exchange interaction energy. Thus, the energy can be written as $\varepsilon_{\bm{k}}=J_0(1-\gamma_{\bm{k}})$, where $\gamma_{\bm{k}}=(1/z)\sum_j J_{0j} \exp{(-i\bm{k} \bm{r}_{0j})}$, $\bm{r}_{0j}=\bm{r}_0-\bm{r}_j$ with
$\bm{r}_{0j}$ the relative position of the $z$ nearest neighbors. 

The integral $\rho(T)=(1/\mathcal{N}) \sum_{\bm{k}} n_{\bm{k}}$ at low temperatures for both quantum and classical statistics are very-well known results.\cite{BastardisPRB2012} 
For the classical statistics
\begin{eqnarray}
m_c(T)&=&
1- \frac{k_{\mathrm{B}} T }{J_0} 
\frac{1}{\mathcal{N}}
\sum_{\bm{k}} \frac{1}{1-\gamma_{\bm{k}}}\approx 
1-\frac{1}{3} \frac{T}{\Tc}, 
\label{eq:classicalSWlowT}
\end{eqnarray}
where $T_c$ is the Curie temperature and we have used the random-phase approximation\cite{GaraninPRB1996} (RPA) relation to relate $W$ and $T_c$ ($J^{\rm{cl}}_0/3\approx W k_B T_c$) (exact for the spherical model \cite{StanleyPR1968}), where  $W=(1/\mathcal{N}\sum_{\bm{k}} \frac{1}{1-\gamma_{\bm{k}}} )$ is the Watson integral. 

Under the same conditions in the quantum Heisenberg case one obtains the $T^{3/2}$ Bloch  law, 
\begin{eqnarray}
m_q(T)&=&
1- \frac{1}{3}s 
 \left(\frac{T}{\Tc}\right)^{3/2}
\label{eq:quantumSWlowT}
\end{eqnarray}
where $s$ is a slope factor given by 
\begin{equation}
s= S^{1/2} \left( 
2\pi W
\right)^{-3/2} \zeta(3/2)\mathrm{.}
\label{eq:coeficientBlochLawCQ}
\end{equation}
where $S$ is the spin integer spin quantum number and $\zeta (x)$ the well-known Riemann $\zeta$ function, 
and the RPA relation for a quantum model ($3 k_{\mathrm{B}} T^{\rm{q}}_c=J^{\rm{q}}_0 S^2/W$) has been used.  
We note that if one wants to have $T^{\rm{q}}_c=T^{\rm{cl}}_c$ then the well-known identification 
$J^{\rm{q}}_0 S^2=J^{\rm{cl}}_0$ is necessary.\cite{BastardisPRB2012}
We also note that Kuz'min\cite{KuzminPRL2005} utilized semi-classical linear spin wave theory to determine $s$, and so use the experimentally measured magnetic moment and avoid the well known problem of choosing a value of $S$ for the studied metals.

Mapping between the classical and quantum $m(T)$ expressions is done simply by equating Eqs.~\eqref{eq:classicalSWlowT} and \eqref{eq:quantumSWlowT} yielding $\tau_{\mathrm{cl}}=s\tau_q^{3/2}$, where $\tau = T/\Tc$, for classical and quantum statistics respectively. This expression therefore relates the thermal fluctuations between the classical and quantum Heisenberg models at low temperatures. At higher temperatures more terms are required to describe $m(T)$ for both approaches, making the simple identification between temperatures cumbersome. At temperatures close to and above $\Tc$,
$\varepsilon_{\bm{k}}/\kB T \rightarrow 0$ is small and thus the thermal Bose distribution $1/(\exp(\varepsilon_{\bm{k}}/\kB T)-1)\approx \varepsilon_{\bm{k}}/\kB T$ tends to the Boltzmann distribution, thus the effect of the spin quantization is negligible here. 
 For this temperature region, a power law is expected, $m(\tau) \approx (1-\tau)^{\beta}$, where $\beta \approx 1/3$ for the Heisenberg model in both cases.

The existence of a simple relation between classical and quantum temperature dependent magnetization at low temperatures leads to the question - does a similar scaling quantitatively describe the behavior of elemental ferromagnets for the whole range of temperatures? Our starting point is to represent the temperature dependent magnetization in the simplest form arising from a straightforward interpolation of the Bloch law\cite{Bloch1930} and critical behavior\cite{chikazumi1997} given by the Curie-Bloch equation
\begin{equation}
m(\tau) = \left(1 - \tau^{\alpha} \right)^{\beta}
\label{eq:mvsT}
\end{equation}
where $\alpha$ is an empirical constant and $\beta \approx 1/3$ is the critical exponent. We will demonstrate that this simple expression is sufficient to describe the temperature dependent magnetization in elemental ferromagnets with a single fitting parameter $\alpha$. An alternative to the Curie-Bloch equation was proposed by Kuz'min\cite{KuzminPRL2005} which has the form
\begin{equation}
m(\tau) = [1-s\tau^{3/2}-(1-s)\tau^p]^{\beta}\mathrm{.}
\label{eq:kuzmin}
\end{equation}
The parameters $s$ and $p$ are taken as fitting parameters, where it was  found that $p=5/2$ for all ferromagnets except for Fe and $s$ relates to the form of the $m(T)$ curve and corresponds to the extent that the magnetization follows Bloch's law at low temperatures. In the case of a pure Bloch ferromagnet where $s=1$, $p = 3/2$ and $\alpha = p$ equations \eqref{eq:mvsT} and \eqref{eq:kuzmin} are identical, demonstrating the same physical origin of these phenomenological equations.

\begin{figure*}[!tb]
\center
\includegraphics[width=17cm, trim=0 0 0 0]{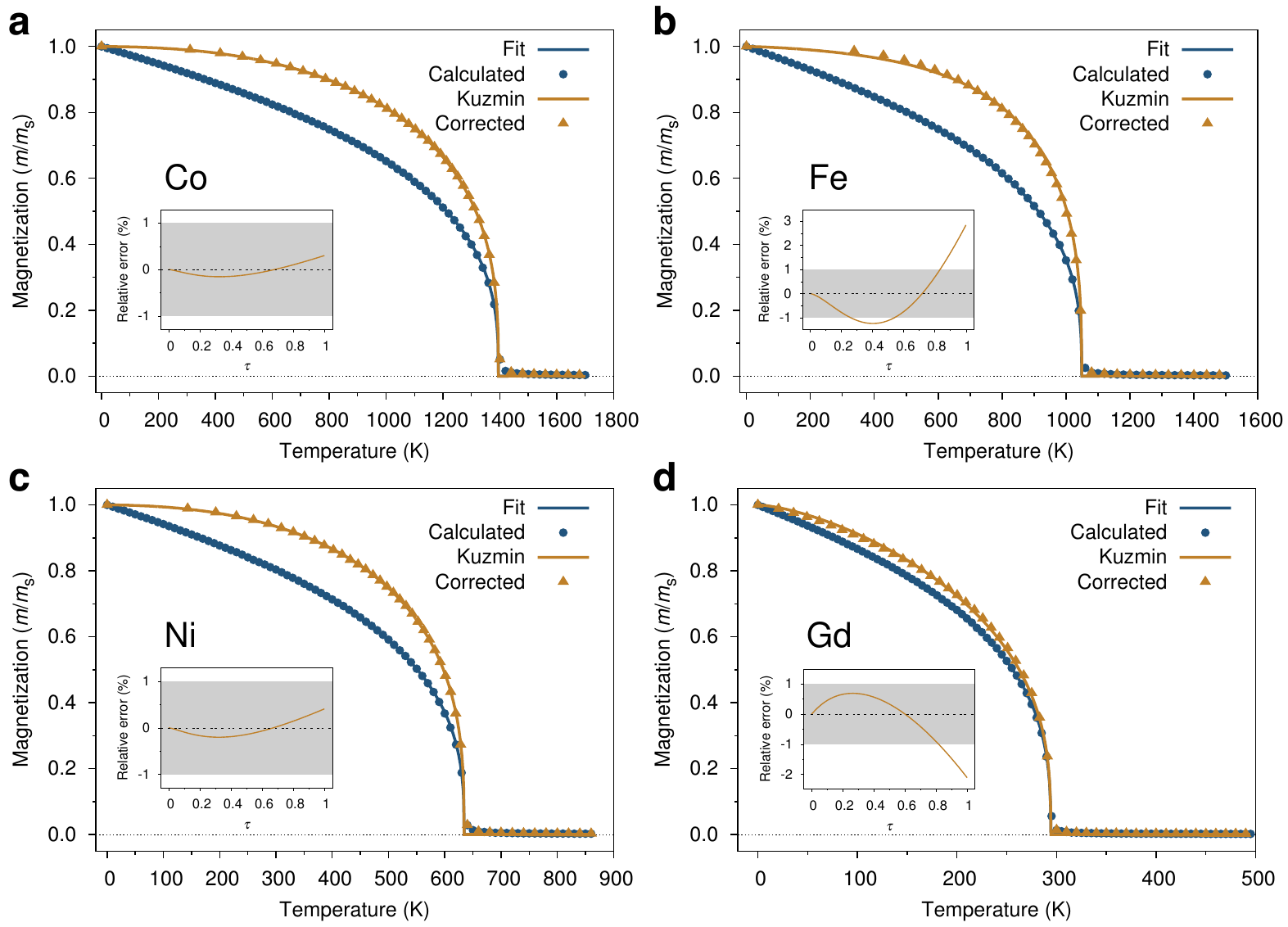}
\caption{Temperature dependent magnetization for the elemental ferromagnets (a) Co, (b) Fe, (c) Ni and (d) Gd. Circles give the simulated mean magnetization, and dark solid lines show the corresponding fit according to Eq.~\eqref{eq:mvsT} for the classical case $\alpha=1$. Light solid lines give the experimentally measured temperature dependent magnetization as fitted by Kuz'min's equation. Triangles give the simulated data after the temperature rescaling has been applied showing excellent agreement with the experimentally measured magnetizations for all studied materials. Inset are plots of the relative error of the rescaled magnetization compared to Kuz'min's fit to the experimental data, showing less than 3\% error for all materials in the whole temperature range (a more restrictive 1\% error is shown by the shaded region). The final fitting parameters are listed in Tab.~\ref{tab:para}. Color Online.}
\label{fig:mvsT}
\end{figure*}

While Kuz'min's equation quantitatively describes the form of the magnetization curve, it does not link the macroscopic Curie temperature to microscopic exchange interactions which can be conveniently determined by \textit{ab-initio} first principles calculations\cite{PajdaBCCFe2001}. Exchange interactions calculated from first principles are often long ranged and oscillatory in nature and so analytical determination of the Curie temperature can be done with a number of different standard approaches such as mean-field (MFA) or random phase approximations (RPA), neither of which are particularly accurate due to the approximations involved. A much more successful method is incorporating the microscopic exchange interactions into a multiscale atomistic spin model which has been shown to yield Curie temperatures much closer to experiment\cite{MyrasovFePtEPL2005}. The clear advantage of this approach is the direct linking of electronic scale calculated parameters to macroscopic thermodynamic magnetic properties such as the Curie temperature. What is interesting is that the classical spin fluctuations give the correct $\Tc$ for a wide range of magnetic materials\cite{PajdaBCCFe2001,MyrasovFePtEPL2005}, suggesting that the particular value of the exchange parameters and the form of the $m(T)$ curve are largely independent quantities.
The difficulty with the classical model is that the form of the curve is intrinsically wrong when compared to experiment.  
\section{Atomistic spin model}
To determine  the classical temperature dependent magnetization for the elemental ferromagnets Co, Fe, Ni and Gd we proceed to simulate them using the classical atomistic spin model. The energetics of the system are described by the classical spin Hamiltonian\cite{EvansVMPR2013} of the form 
\begin{equation}
\mathscr{H} = -\sum_{i<j} \smJij \sms_i \cdot \sms_j
\label{eqn:ham}
\end{equation}
where $\sms_i$ and $\sms_j$ are unit vectors describing the direction of the local and nearest neighbor magnetic moments at each atomic site and \smJij is the nearest neighbor exchange energy given by\cite{GaraninPRB1996}
\begin{equation}
\smJij = \frac{3 k_B \Tc}{\gamma z}
\label{eq:JijTc}
\end{equation}
where $\gamma (W)$ gives a correction factor from the MFA and which for RPA $\gamma=1/W$ and the value of \Tc is taken from experiment. The numerical calculations have been carried out using the \vampire software package\cite{vampire-url}. The simulated system for Co, Ni, Fe and Gd consists of a cube (20 nm)$^3$ in size with periodic boundary conditions applied to reduce finite-size effects by eliminating the surface. The equilibrium temperature dependent properties of the system are calculated using the Hinzke-Nowak Monte Carlo algorithm\cite{EvansVMPR2013,HinzkeMC1999} using 20,000 equilibration steps and 20,000 averaging steps resulting in the calculated temperature dependent magnetization curves for each element shown in Fig.~\ref{fig:mvsT}. For a classical spin model it is known that the simulated temperature dependent magnetization is well fitted by the function\cite{EvansVMPR2013}
\begin{equation}
m(T) = \left(1 - \frac{T}{\Tc}\right)^{\beta}\mathrm{.}
\label{eq:cmvsT}
\end{equation}
We note that Eqs.~\ref{eq:mvsT} and \ref{eq:cmvsT} are identical for the case of $\alpha = 1$. Fitting the simulated temperature dependent magnetization for Fe, Co, Ni and Gd to Eq.~\ref{eq:cmvsT} in our case yields an apparently universal critical exponent of $\beta = 0.340 \pm 0.001$ and a good estimate of the Curie temperature, $\Tc$ within 1\% of the experimental values. In general $\beta$ depends on both the system size and on the form of the spin Hamiltonian\cite{hovorkaAPLTcDist2012}, hence our use of a large system size and many averaging Monte Carlo steps. We note that our calculated critical exponent in all cases is closer to 0.34 as found experimentally for Ni\cite{Anderson1971} rather than the $^1/_3$ normally expected.\cite{KuzminPRL2005} The simulations confirm the ability of the atomistic spin model to relate microscopic exchange interactions to the macroscopic Curie temperature. However as is evident from the Kuz'min fits to the experimental data (see Fig.~\ref{fig:mvsT}) the form of the magnetization curve is seriously in error. 
\section{Temperature rescaling}
\begin{figure}[!t]
\center
\includegraphics[width=7cm, trim=0 0 0 0]{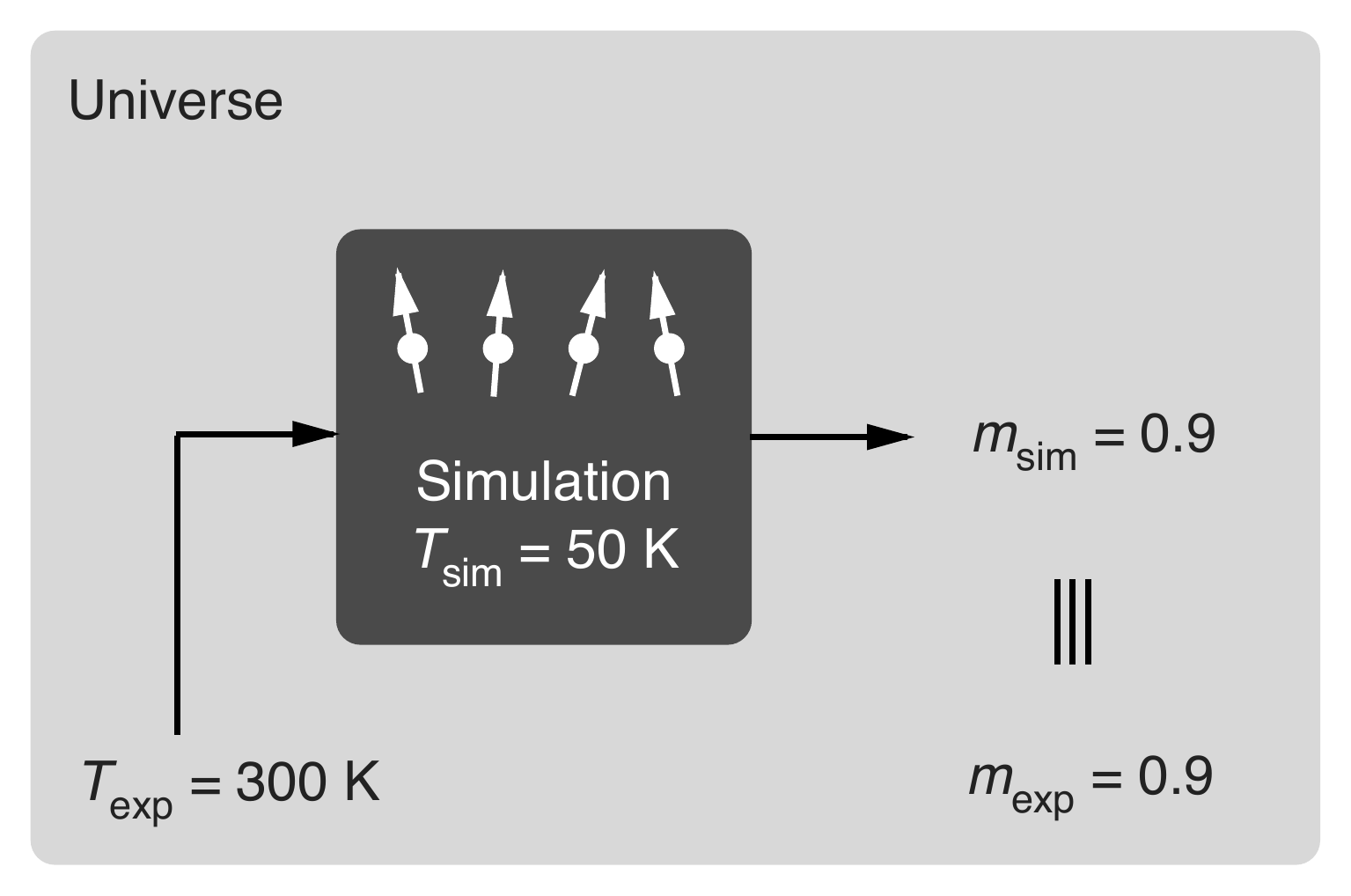}
\caption{Schematic diagram of the rescaling applied to the simulation of a magnetic material. The universe has a temperature $\Texp = 300$K, which for an experimental sample has a macroscopic magnetization length of $m_{\mathrm{exp}} = M/M_{\mathrm{s}}^0$ = 0.9. Using the temperature rescaling this leads to an internal simulation temperature of $\Tsim = 50$K, which leads to a simulated equilibrium magnetization of $m_{\mathrm{sim}} = 0.9$. Therefore macroscopically $m_{\mathrm{exp}} \equiv m_{\mathrm{sim}}$.} 
\label{fig:schematic}
\end{figure}

To resolve the disparity in the temperature dependent magnetization between the classical simulation and experiment we proceed by implementing temperature rescaling to map the simulations onto experiment in a quantitative manner. Similar to Kuz'min\cite{KuzminPRL2005}, we assume in our fitting that the critical exponent $\beta$ is universal and thus the same for both the classical simulation and for experiment, and so the only free fitting parameter is $\alpha$. Due to the limited availability of raw experimental data, we use the equation proposed by Kuz'min as a substitute for the experimental data, since they agree extremely well.\cite{KuzminPRL2005} This also has the advantage of smoothing any errors in experimental data. We proceed by fitting the Curie-Bloch equation given by Eq.~\ref{eq:mvsT} to the Kuz'min equation given by Eq.~\ref{eq:kuzmin} where the parameters $s$ and $p$ are known fitting parameters (determined from experimental data by Kuz'min\cite{KuzminPRL2005}), and $\beta \simeq 0.34$ and $\Tc$ are determined from the atomistic simulations. The determined value of $\alpha$ then conveniently relates the result of the classical simulation to the experimental data, allowing a simple mapping as follows. The (internal) simulation temperature \Tsim is rescaled so that for the input experimental (external) temperature \Texp the equilibrium magnetization agrees with the experimental result. \Tsim and \Texp are related by the expression
\begin{equation}
\frac{\Tsim}{\Tc} = \left(\frac{\Texp}{\Tc}\right)^{\alpha}\mathrm{.}
\label{eq:vT}
\end{equation}
Thus, for a desired \textit{real} temperature \Texp, the simulation will use an effective temperature within the Monte Carlo or Langevin dynamics simulation of \Texp, which for $\alpha >1$, $\Tsim < \Texp$ leading to an effective reduction of the thermal fluctuations in the simulation. The physical interpretation of the rescaling is that at low temperatures the allowed spin fluctuations in the classical limit are over estimated and so this corresponds to a higher effective temperature than given in the simulation. This is illustrated schematically in Fig.~\ref{fig:schematic}.

\begin{figure}[!t]
\center
\includegraphics[width=8cm, trim=0 0 0 0]{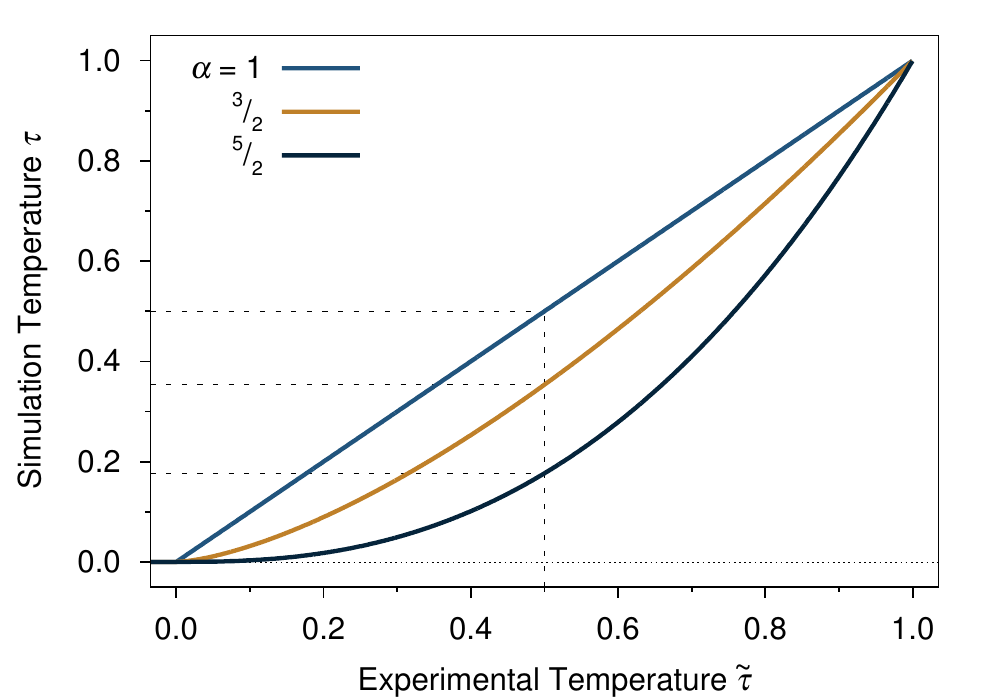}
\caption{Plot of reduced simulation temperature $\tau = \Tsim/\Tc$ as a function of the reduced input experimental temperature $\widetilde{\tau} = \Texp/\Tc$ for different values of the rescaling exponent $\alpha$. Higher values of $\alpha$ correspond to a lower effective temperature and reduced fluctuations in the simulation.}
\label{fig:rT}
\end{figure}

Clearly different values of $\alpha$ in Eq.~\ref{eq:vT} lead to different mappings between the experimental temperature and the internal simulation temperature. Larger values of $\alpha$ lead to reduced thermal fluctuations in the spin model simulations, owing to quantum mechanical ``stiffness''. A plot of the simulation temperature \Tsim as a function of the input experimental temperature \Texp for different values of the rescaling exponent $\alpha$ is shown in Fig.~\ref{fig:rT}. Above \Tc it is assumed that $\Tsim = \Texp$ due to the absence of magnetic order.
For Monte Carlo simulations the reduced simulation temperature appears directly in the acceptance criteria $P = \exp\left(-\Delta E / \kB \Tsim \right)$ for individual trial moves, thus reducing the probability of acceptance and resulting in a larger magnetization length for the system.

\begin{table}[!t]
\caption{Fitting parameters for the temperature dependent magnetization derived from the classical spin model simulations by fitting to Eq.~\eqref{eq:mvsT} for $\alpha = 1$ ($\Tc$ and $\beta$) and by secondary fitting to Eq.~\eqref{eq:kuzmin} to obtain the rescaling factor $\alpha$.}
\begin{ruledtabular}
\begin{tabular}{ l c c c c }
  & Co & Fe & Ni & Gd\\
\hline
\Tc        & 1395 K  & 1049 K  & 635 K   & 294 K\\
$\beta$    & 0.340 & 0.339 & 0.341 & 0.339\\
$\alpha$   & 2.369 & 2.876 & 2.322 & 1.278\\
\end{tabular}
\end{ruledtabular}
\label{tab:para}
\end{table}

We now apply the temperature rescaling to the simulated temperature dependent magnetization for Fe, Co, Ni and Gd and directly compare to the experimental curve, as shown by the corrected simulation data in Fig.~\ref{fig:mvsT}, where the final fitted parameters are given in Tab.~\ref{tab:para}. For Co, Ni and Gd the agreement between the rescaled simulation data and the experimental measurement is remarkable given the simplicity of the approach. The fit for Fe is not as good as for the others due to the peculiarity of the experimentally measured magnetization curve, as noted by Kuz'min\cite{KuzminPRL2005}. However the simple rescaling presented here is accurate to a few percent over the whole temperature range, and if greater accuracy is required then a non-analytic temperature rescaling can be used to give exact agreement with the experimental data.

The ability of direct interpolation of Bloch's Law with critical scaling to describe the temperature dependent magnetization is significant for two reasons. Firstly, it provides a simple way to parameterize experimentally measured temperature dependent magnetization in terms of only three parameters via Eq.~\eqref{eq:mvsT}. Secondly, it allows a direct and more accurate determination of the temperature dependence of all the parameters needed for numerical micromagnetics at elevated temperatures from first principles when combined with atomistic spin model simulations\cite{KazantsevaPRB2008,AtxitiaEXC2010,AsselinCMC2010}. We also expect the same form is applicable to other technologically important composite magnets such as CoFeB, NdFeB or FePt alloys. 

\section{Dynamic temperature rescaling}
We now proceed to demonstrate the power of the rescaling method by considering magnetization dynamics using a Langevin dynamics approach\cite{EvansVMPR2013} with temperature rescaling. The temperature rescaling can be used for equilibrium simulations at constant temperature, but also dynamic simulations where the temperature changes continuously. The latter is particularly important for simulating the effects of laser heating and also spin caloritronics with dynamic heating. As an example, we simulate the laser-induced sub picosecond demagnetization of Ni first observed experimentally by Beaurepaire \textit{et al.} \cite{BeaurepairePRL1996}. The energetics of our Ni model are given by the Heisenberg spin Hamiltonian
\begin{equation}
\HH = -\sum_{i<j} \smJij \sms_i \cdot \sms_j -\sum_{i} \smku S_{i,z}^2
\label{Hamiltonian}
\end{equation}
where $\smJij = 2.757 \times 10^{-21}$ J/link is the exchange energy between nearest neighboring Ni spins, $\sms_i$ and $\sms_j$ are a unit vectors describing the direction of the local and neighboring spin moments respectively and $\smku = 5.47\times 10^{-26}$ J/atom.

The dynamics of each atomic spin is given by the stochastic Landau-Lifshitz-Gilbert (sLLG) equation applied at the atomistic level given by
\begin{equation}\label{eqn:LLG}
\frac{\partial \sms_i }{\partial t} =
-\frac{\gamma_e}{(1+\lambda^{2})}[\sms_i \times
\smH^i_{\mathrm{eff}} + \lambda \sms_i \times
(\sms_i \times \smH^i_{\mathrm{eff}})]
\end{equation}
where $\gamma_e=1.76 \times 10^{11}$ JT$^{-1}$s$^{-1}$ is the gyromagnetic ratio, $\lambda = 0.001$ is the phenomenological Gilbert damping parameter, and $\smH^i_{\mathrm{eff}}$ is the net magnetic field on each atomic spin. The sLLG equation describes the interaction of an atomic spin moment $i$ with an effective magnetic field, which is obtained from the derivative of the spin Hamiltonian and the addition of a Langevin thermal term, giving a total effective field on each spin
\begin{equation}\label{eqn:Heffth}
  \smH^i_{\mathrm{eff}} = -\frac{1}{\smmu}\frac{\partial \HH}{\partial \sms_i} + \smH^{i}_{\mathrm{th}}
\end{equation}
where $\smmu = 0.606 \muB$ is the atomic spin moment. The thermal field in each spatial dimension is represented by a normal distribution $\boldsymbol{\Gamma}(t)$ with a standard deviation of 1 and mean of zero. The thermal field is given by
\begin{equation}
\smH^{i}_{\mathrm{th}} = \boldsymbol{\Gamma}(t) \sqrt{\frac{2 \lambda \kB \Tsim }{\gamma_e \smmu \Delta t}}
\end{equation}
where  $k_{\mathrm{B}}$ is the Boltzmann constant, $\Delta t$ is the integration time step and $\Tsim$ is the rescaled simulation temperature from Eq.~\ref{eq:vT}. As with the Monte Carlo simulations, this reduces the thermal fluctuations in the sLLG and leads to higher equilibrium magnetization length compared to usual classical simulations. However unlike Monte Carlo simulations, the explicit timescale in the sLLG equation allows the simulation of dynamic processes, particularly with dynamic changes in the temperature associated with ultrafast laser heating. In this case the temporal evolution of the electron temperature can be calculated using a \textit{two temperature model}\cite{AnisimovTTM1974}, considering the dynamic response of the electron ($T_e^{\mathrm{exp}}$) and lattice ($T_l^{\mathrm{exp}}$) temperatures. To be explicit, when including the temperature rescaling the two temperature model always refers to the real, or experimental temperature, \Texp; \Tsim only applies to the magnetic part of the simulation where the thermal fluctuations are included. The time evolution of $T_e^{\mathrm{exp}}$ and $T_l^{\mathrm{exp}}$ is given by\cite{AnisimovTTM1974}
\begin{eqnarray}
C_e\frac{\partial T_e^{\mathrm{exp}}}{\partial t} &=& -G(T_e^{\mathrm{exp}}-T_l^{\mathrm{exp}})+S(t) \\
C_l\frac{\partial T_l^{\mathrm{exp}}}{\partial t} &=& -G(T_l^{\mathrm{exp}}-T_e^{\mathrm{exp}})
\end{eqnarray}
where $C_e$ and $C_l$ are the electron and lattice heat capacities, $G$ is the electron-lattice coupling factor, and $S(t)$ is a time-dependent Gaussian pulse with a FWHM of 60 fs which adds energy to the electron system representing the laser pulse. The time evolution of the electron temperature is solved numerically using a simple Euler scheme. The parameters used are representative of Ni\cite{LinPRB2008}, with $G = 12 \times 10^{17}$ W m$^{-3}$K$^{-1}$, $C_e = 8 \times 10^{2}$ J m$^{-3}$K$^{-1}$ and $C_l = 4 \times 10^{6}$ J m$^{-3}$K$^{-1}$. The sLLG is solved numerically using the time dependent electron temperature rescaled using Eq.~\ref{eq:vT} with the Heun numerical scheme\cite{EvansVMPR2013} and a timestep of $\Delta t = 1 \times 10^{-16}$ s.

\begin{figure}[!t]
\center
\includegraphics[width=8cm, trim=0 0 0 0]{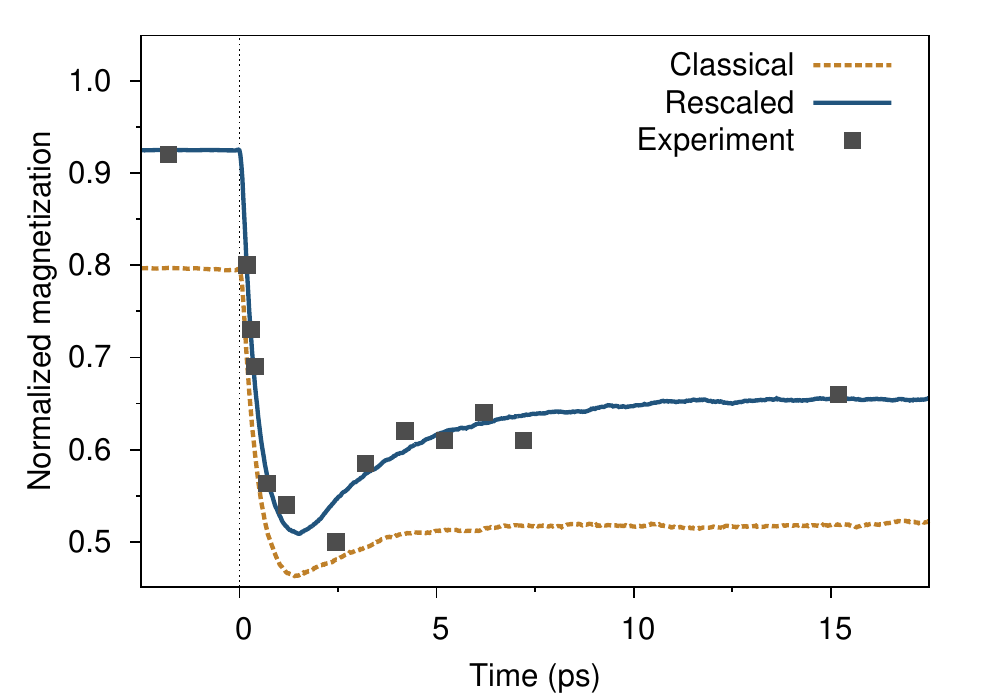}
\caption{Simulated demagnetization of Ni comparing classical and rescaled models with experimental data from [\onlinecite{BeaurepairePRL1996}]. The rescaled dynamic simulations show quantitative agreement with experiment from an atomic level model. Color Online.}
\label{fig:demag}
\end{figure}
To simulate the effects of a laser pulse on Ni, we model a small system of (8 nm)$^3$ which is first equilibrated at \Texp = 300 K for 20ps, sufficient to thermalize the system. The temperature of the spin system is linked to the electron temperature and so a simulated laser pulse leads to a transient increase of the temperature inducing ultrafast magnetization dynamics. After a few ps the energy is transferred to the lattice where $T_e^{\mathrm{exp}} = T_l^{\mathrm{exp}}$. The classical and rescaled dynamics are calculated for identical parameters except that $\alpha$ = 1 is used for the classical simulation since no rescaling is used. The simulated magnetization dynamics alongside the experimental results are shown in Fig.~\ref{fig:demag}, where the laser pulse arrives at $t=0$. As expected the standard classical model shows poor agreement with experiment because of the incorrect $m(T)$. However, after applying dynamic temperature rescaling \textit{quantitative} agreement is found between the atomistic model and experiment. This result exemplifies the validity of our approach by demonstrating the ability to describe both equilibrium and dynamic properties of magnetic materials at all temperatures.
\section{Discussion and conclusion}
In conclusion, we have performed atomistic spin model simulations of the temperature dependent magnetization of the elemental ferromagnets Ni, Fe, Co and Gd to determine the Curie temperature directly from the microscopic exchange interactions. Using a simple temperature rescaling considering classical and quantum spin wave fluctuations we find quantitative agreement between the simulations and experiment for the temperature dependent magnetization. By rescaling the temperature in this way it is now possible to derive all temperature dependent magnetic properties in quantitative agreement with experiment from a microscopic atomistic model. In addition we have shown the applicability of the approach to modeling ultrafast magnetization dynamics, also in quantitative agreement with experiment. This approach now enables accurate temperature dependent simulations of magnetic materials suitable for a wide range of materials of practical and fundamental interest.

Finally it is interesting to ponder what is the physical origin of the exponent $\alpha$. From the elements studied in this paper, there is no correlation between $\alpha$ and the crystallographic structure or the Curie temperature, nor by extension the strength of the interatomic exchange constant. The rescaling is independent of temperature and so the origin must be an intrinsic property of the system with a quantum mechanical origin as suggested by Eq. \eqref{eq:coeficientBlochLawCQ}. In the simplistic picture it should relate to the availability of spin states in the vicinity of the ground state, with the fewer available states the more Bloch-like the temperature dependent magnetization will be. However, it would be interesting to apply detailed \emph{ab-initio} calculations to try and delineate the origin of this effect in simple ferromagnets.

\section{Acknowledgements}
This work was supported by the European Community's Seventh Framework Programme (FP7/2007-2013) under  Grant Agreement No. 281043 \textsc{femtospin}. UA gratefully acknowledges support from  Basque Country Government under "Programa Posdoctoral de perfeccionamiento de doctores del DEUI del Gobierno Vasco" and  EU FP7 Marie Curie Zukunftskolleg Incoming Fellowship Programme (Grant No. 291784), University of Konstanz. 

\bibliography{library,local}

\begin{thebibliography}{36}%
\makeatletter
\providecommand \@ifxundefined [1]{%
 \@ifx{#1\undefined}
}%
\providecommand \@ifnum [1]{%
 \ifnum #1\expandafter \@firstoftwo
 \else \expandafter \@secondoftwo
 \fi
}%
\providecommand \@ifx [1]{%
 \ifx #1\expandafter \@firstoftwo
 \else \expandafter \@secondoftwo
 \fi
}%
\providecommand \natexlab [1]{#1}%
\providecommand \enquote  [1]{``#1''}%
\providecommand \bibnamefont  [1]{#1}%
\providecommand \bibfnamefont [1]{#1}%
\providecommand \citenamefont [1]{#1}%
\providecommand \href@noop [0]{\@secondoftwo}%
\providecommand \href [0]{\begingroup \@sanitize@url \@href}%
\providecommand \@href[1]{\@@startlink{#1}\@@href}%
\providecommand \@@href[1]{\endgroup#1\@@endlink}%
\providecommand \@sanitize@url [0]{\catcode `\\12\catcode `\$12\catcode
  `\&12\catcode `\#12\catcode `\^12\catcode `\_12\catcode `\%12\relax}%
\providecommand \@@startlink[1]{}%
\providecommand \@@endlink[0]{}%
\providecommand \url  [0]{\begingroup\@sanitize@url \@url }%
\providecommand \@url [1]{\endgroup\@href {#1}{\urlprefix }}%
\providecommand \urlprefix  [0]{URL }%
\providecommand \Eprint [0]{\href }%
\providecommand \doibase [0]{http://dx.doi.org/}%
\providecommand \selectlanguage [0]{\@gobble}%
\providecommand \bibinfo  [0]{\@secondoftwo}%
\providecommand \bibfield  [0]{\@secondoftwo}%
\providecommand \translation [1]{[#1]}%
\providecommand \BibitemOpen [0]{}%
\providecommand \bibitemStop [0]{}%
\providecommand \bibitemNoStop [0]{.\EOS\space}%
\providecommand \EOS [0]{\spacefactor3000\relax}%
\providecommand \BibitemShut  [1]{\csname bibitem#1\endcsname}%
\let\auto@bib@innerbib\@empty
\bibitem [{\citenamefont {Gutfleisch}\ \emph {et~al.}(2011)\citenamefont
  {Gutfleisch}, \citenamefont {Willard}, \citenamefont {Brück}, \citenamefont
  {Chen}, \citenamefont {Sankar},\ and\ \citenamefont
  {Liu}}]{GutfleischAM2011}%
  \BibitemOpen
  \bibfield  {author} {\bibinfo {author} {\bibfnamefont {O.}~\bibnamefont
  {Gutfleisch}}, \bibinfo {author} {\bibfnamefont {M.~A.}\ \bibnamefont
  {Willard}}, \bibinfo {author} {\bibfnamefont {E.}~\bibnamefont {Brück}},
  \bibinfo {author} {\bibfnamefont {C.~H.}\ \bibnamefont {Chen}}, \bibinfo
  {author} {\bibfnamefont {S.~G.}\ \bibnamefont {Sankar}}, \ and\ \bibinfo
  {author} {\bibfnamefont {J.~P.}\ \bibnamefont {Liu}},\ }\href {\doibase
  10.1002/adma.201002180} {\bibfield  {journal} {\bibinfo  {journal} {Adv.
  Mater.}\ }\textbf {\bibinfo {volume} {23}},\ \bibinfo {pages} {821} (\bibinfo
  {year} {2011})}\BibitemShut {NoStop}%
\bibitem [{\citenamefont {M.~Plumer}\ and\ \citenamefont
  {Weller}(2001)}]{Plumer2001}%
  \BibitemOpen
  \bibfield  {author} {\bibinfo {author} {\bibfnamefont {J.~v.~E.}\
  \bibnamefont {M.~Plumer}}\ and\ \bibinfo {author} {\bibfnamefont
  {D.}~\bibnamefont {Weller}},\ }\href
  {http://www.springer.com/materials/book/978-3-540-42370-6} {\emph {\bibinfo
  {title} {The Physics of Ultra-High-Density Magnetic Recording}}},\ Springer
  Series in Surface Sciences, Vol. 41\ (\bibinfo {year} {2001})\BibitemShut
  {NoStop}%
\bibitem [{\citenamefont {Parkin}\ \emph {et~al.}(2008)\citenamefont {Parkin},
  \citenamefont {Hayashi},\ and\ \citenamefont {Thomas}}]{ParkinScience2008}%
  \BibitemOpen
  \bibfield  {author} {\bibinfo {author} {\bibfnamefont {S.~S.~P.}\
  \bibnamefont {Parkin}}, \bibinfo {author} {\bibfnamefont {M.}~\bibnamefont
  {Hayashi}}, \ and\ \bibinfo {author} {\bibfnamefont {L.}~\bibnamefont
  {Thomas}},\ }\href {\doibase 10.1126/science.1145799} {\bibfield  {journal}
  {\bibinfo  {journal} {Science}\ }\textbf {\bibinfo {volume} {320}},\ \bibinfo
  {pages} {190} (\bibinfo {year} {2008})}\BibitemShut {NoStop}%
\bibitem [{\citenamefont {Wolf}\ \emph {et~al.}(2001)\citenamefont {Wolf},
  \citenamefont {Awschalom}, \citenamefont {Buhrman}, \citenamefont {Daughton},
  \citenamefont {von Molnár}, \citenamefont {Roukes}, \citenamefont
  {Chtchelkanova},\ and\ \citenamefont {Treger}}]{WolfScience2001}%
  \BibitemOpen
  \bibfield  {author} {\bibinfo {author} {\bibfnamefont {S.~A.}\ \bibnamefont
  {Wolf}}, \bibinfo {author} {\bibfnamefont {D.~D.}\ \bibnamefont {Awschalom}},
  \bibinfo {author} {\bibfnamefont {R.~A.}\ \bibnamefont {Buhrman}}, \bibinfo
  {author} {\bibfnamefont {J.~M.}\ \bibnamefont {Daughton}}, \bibinfo {author}
  {\bibfnamefont {S.}~\bibnamefont {von Molnár}}, \bibinfo {author}
  {\bibfnamefont {M.~L.}\ \bibnamefont {Roukes}}, \bibinfo {author}
  {\bibfnamefont {A.~Y.}\ \bibnamefont {Chtchelkanova}}, \ and\ \bibinfo
  {author} {\bibfnamefont {D.~M.}\ \bibnamefont {Treger}},\ }\href {\doibase
  10.1126/science.1065389} {\bibfield  {journal} {\bibinfo  {journal}
  {Science}\ }\textbf {\bibinfo {volume} {294}},\ \bibinfo {pages} {1488}
  (\bibinfo {year} {2001})}\BibitemShut {NoStop}%
\bibitem [{\citenamefont {Jordan}\ \emph {et~al.}(1999)\citenamefont {Jordan},
  \citenamefont {Scholz}, \citenamefont {Wust}, \citenamefont {Fähling},\ and\
  \citenamefont {Felix}}]{JordanJMMM1999}%
  \BibitemOpen
  \bibfield  {author} {\bibinfo {author} {\bibfnamefont {A.}~\bibnamefont
  {Jordan}}, \bibinfo {author} {\bibfnamefont {R.}~\bibnamefont {Scholz}},
  \bibinfo {author} {\bibfnamefont {P.}~\bibnamefont {Wust}}, \bibinfo {author}
  {\bibfnamefont {H.}~\bibnamefont {Fähling}}, \ and\ \bibinfo {author}
  {\bibfnamefont {R.}~\bibnamefont {Felix}},\ }\href {\doibase
  http://dx.doi.org/10.1016/S0304-8853(99)00088-8} {\bibfield  {journal}
  {\bibinfo  {journal} {J. Magn. Magn. Mater.}\ }\textbf {\bibinfo {volume}
  {201}},\ \bibinfo {pages} {413 } (\bibinfo {year} {1999})}\BibitemShut
  {NoStop}%
\bibitem [{\citenamefont {Beaurepaire}\ \emph {et~al.}(1996)\citenamefont
  {Beaurepaire}, \citenamefont {Merle}, \citenamefont {Daunois},\ and\
  \citenamefont {Bigot}}]{BeaurepairePRL1996}%
  \BibitemOpen
  \bibfield  {author} {\bibinfo {author} {\bibfnamefont {E.}~\bibnamefont
  {Beaurepaire}}, \bibinfo {author} {\bibfnamefont {J.-C.}\ \bibnamefont
  {Merle}}, \bibinfo {author} {\bibfnamefont {A.}~\bibnamefont {Daunois}}, \
  and\ \bibinfo {author} {\bibfnamefont {J.-Y.}\ \bibnamefont {Bigot}},\ }\href
  {\doibase 10.1103/PhysRevLett.76.4250} {\bibfield  {journal} {\bibinfo
  {journal} {Phys. Rev. Lett.}\ }\textbf {\bibinfo {volume} {76}},\ \bibinfo
  {pages} {4250} (\bibinfo {year} {1996})}\BibitemShut {NoStop}%
\bibitem [{\citenamefont {Radu}\ \emph {et~al.}(2011)\citenamefont {Radu},
  \citenamefont {Vahaplar}, \citenamefont {Stamm}, \citenamefont {Kachel},
  \citenamefont {Pontius}, \citenamefont {Durr}, \citenamefont {Ostler},
  \citenamefont {Barker}, \citenamefont {Evans}, \citenamefont {Chantrell},
  \citenamefont {Tsukamoto}, \citenamefont {Itoh}, \citenamefont {Kirilyuk},
  \citenamefont {Rasing},\ and\ \citenamefont {Kimel}}]{RaduNature2011}%
  \BibitemOpen
  \bibfield  {author} {\bibinfo {author} {\bibfnamefont {I.}~\bibnamefont
  {Radu}}, \bibinfo {author} {\bibfnamefont {K.}~\bibnamefont {Vahaplar}},
  \bibinfo {author} {\bibfnamefont {C.}~\bibnamefont {Stamm}}, \bibinfo
  {author} {\bibfnamefont {T.}~\bibnamefont {Kachel}}, \bibinfo {author}
  {\bibfnamefont {N.}~\bibnamefont {Pontius}}, \bibinfo {author} {\bibfnamefont
  {H.~A.}\ \bibnamefont {Durr}}, \bibinfo {author} {\bibfnamefont {T.~A.}\
  \bibnamefont {Ostler}}, \bibinfo {author} {\bibfnamefont {J.}~\bibnamefont
  {Barker}}, \bibinfo {author} {\bibfnamefont {R.~F.~L.}\ \bibnamefont
  {Evans}}, \bibinfo {author} {\bibfnamefont {R.~W.}\ \bibnamefont
  {Chantrell}}, \bibinfo {author} {\bibfnamefont {A.}~\bibnamefont
  {Tsukamoto}}, \bibinfo {author} {\bibfnamefont {A.}~\bibnamefont {Itoh}},
  \bibinfo {author} {\bibfnamefont {A.}~\bibnamefont {Kirilyuk}}, \bibinfo
  {author} {\bibfnamefont {T.}~\bibnamefont {Rasing}}, \ and\ \bibinfo {author}
  {\bibfnamefont {A.~V.}\ \bibnamefont {Kimel}},\ }\href {\doibase
  10.1038/nature09901} {\bibfield  {journal} {\bibinfo  {journal} {Nature}\
  }\textbf {\bibinfo {volume} {472}},\ \bibinfo {pages} {205} (\bibinfo {year}
  {2011})}\BibitemShut {NoStop}%
\bibitem [{\citenamefont {Ostler}\ \emph {et~al.}(2012)\citenamefont {Ostler},
  \citenamefont {Barker}, \citenamefont {Evans}, \citenamefont {Chantrell},
  \citenamefont {Atxitia}, \citenamefont {Chubykalo-Fesenko}, \citenamefont
  {El~Moussaoui}, \citenamefont {Le~Guyader}, \citenamefont {Mengotti},
  \citenamefont {Heyderman}, \citenamefont {Nolting}, \citenamefont
  {Tsukamoto}, \citenamefont {Itoh}, \citenamefont {Afanasiev}, \citenamefont
  {Ivanov}, \citenamefont {Kalashnikova}, \citenamefont {Vahaplar},
  \citenamefont {Mentink}, \citenamefont {Kirilyuk}, \citenamefont {Rasing},\
  and\ \citenamefont {Kimel}}]{OstlerNatCom2012}%
  \BibitemOpen
  \bibfield  {author} {\bibinfo {author} {\bibfnamefont {T.~A.}\ \bibnamefont
  {Ostler}}, \bibinfo {author} {\bibfnamefont {J.}~\bibnamefont {Barker}},
  \bibinfo {author} {\bibfnamefont {R.~F.~L.}\ \bibnamefont {Evans}}, \bibinfo
  {author} {\bibfnamefont {R.~W.}\ \bibnamefont {Chantrell}}, \bibinfo {author}
  {\bibfnamefont {U.}~\bibnamefont {Atxitia}}, \bibinfo {author} {\bibfnamefont
  {O.}~\bibnamefont {Chubykalo-Fesenko}}, \bibinfo {author} {\bibfnamefont
  {S.}~\bibnamefont {El~Moussaoui}}, \bibinfo {author} {\bibfnamefont
  {L.}~\bibnamefont {Le~Guyader}}, \bibinfo {author} {\bibfnamefont
  {E.}~\bibnamefont {Mengotti}}, \bibinfo {author} {\bibfnamefont {L.~J.}\
  \bibnamefont {Heyderman}}, \bibinfo {author} {\bibfnamefont {F.}~\bibnamefont
  {Nolting}}, \bibinfo {author} {\bibfnamefont {A.}~\bibnamefont {Tsukamoto}},
  \bibinfo {author} {\bibfnamefont {A.}~\bibnamefont {Itoh}}, \bibinfo {author}
  {\bibfnamefont {D.}~\bibnamefont {Afanasiev}}, \bibinfo {author}
  {\bibfnamefont {B.~A.}\ \bibnamefont {Ivanov}}, \bibinfo {author}
  {\bibfnamefont {A.~M.}\ \bibnamefont {Kalashnikova}}, \bibinfo {author}
  {\bibfnamefont {K.}~\bibnamefont {Vahaplar}}, \bibinfo {author}
  {\bibfnamefont {J.}~\bibnamefont {Mentink}}, \bibinfo {author} {\bibfnamefont
  {A.}~\bibnamefont {Kirilyuk}}, \bibinfo {author} {\bibfnamefont
  {T.}~\bibnamefont {Rasing}}, \ and\ \bibinfo {author} {\bibfnamefont {A.~V.}\
  \bibnamefont {Kimel}},\ }\href {\doibase 10.1038/ncomms1666} {\bibfield
  {journal} {\bibinfo  {journal} {Nat. Commun.}\ }\textbf {\bibinfo {volume}
  {3}},\ \bibinfo {pages} {666} (\bibinfo {year} {2012})}\BibitemShut {NoStop}%
\bibitem [{\citenamefont {Bauer}\ \emph {et~al.}(2012)\citenamefont {Bauer},
  \citenamefont {Saitoh},\ and\ \citenamefont {van Wees}}]{BauerNMat2012}%
  \BibitemOpen
  \bibfield  {author} {\bibinfo {author} {\bibfnamefont {G.~E.~W.}\
  \bibnamefont {Bauer}}, \bibinfo {author} {\bibfnamefont {E.}~\bibnamefont
  {Saitoh}}, \ and\ \bibinfo {author} {\bibfnamefont {B.~J.}\ \bibnamefont {van
  Wees}},\ }\href {\doibase 10.1038/nmat3301} {\bibfield  {journal} {\bibinfo
  {journal} {Nat. Mater.}\ }\textbf {\bibinfo {volume} {11}},\ \bibinfo {pages}
  {391} (\bibinfo {year} {2012})}\BibitemShut {NoStop}%
\bibitem [{\citenamefont {Kryder}\ \emph {et~al.}(2008)\citenamefont {Kryder},
  \citenamefont {Gage}, \citenamefont {McDaniel}, \citenamefont {Challener},
  \citenamefont {Rottmayer}, \citenamefont {Ju}, \citenamefont {Hsia},\ and\
  \citenamefont {Erden}}]{KryderIEEE2008}%
  \BibitemOpen
  \bibfield  {author} {\bibinfo {author} {\bibfnamefont {M.}~\bibnamefont
  {Kryder}}, \bibinfo {author} {\bibfnamefont {E.}~\bibnamefont {Gage}},
  \bibinfo {author} {\bibfnamefont {T.}~\bibnamefont {McDaniel}}, \bibinfo
  {author} {\bibfnamefont {W.}~\bibnamefont {Challener}}, \bibinfo {author}
  {\bibfnamefont {R.}~\bibnamefont {Rottmayer}}, \bibinfo {author}
  {\bibfnamefont {G.}~\bibnamefont {Ju}}, \bibinfo {author} {\bibfnamefont
  {Y.-T.}\ \bibnamefont {Hsia}}, \ and\ \bibinfo {author} {\bibfnamefont
  {M.}~\bibnamefont {Erden}},\ }\href {\doibase 10.1109/JPROC.2008.2004315}
  {\bibfield  {journal} {\bibinfo  {journal} {Proceedings of the IEEE}\
  }\textbf {\bibinfo {volume} {96}},\ \bibinfo {pages} {1810} (\bibinfo {year}
  {2008})}\BibitemShut {NoStop}%
\bibitem [{\citenamefont {Prejbeanu}\ \emph {et~al.}(2007)\citenamefont
  {Prejbeanu}, \citenamefont {Kerekes}, \citenamefont {Sousa}, \citenamefont
  {Sibuet}, \citenamefont {Redon}, \citenamefont {Dieny},\ and\ \citenamefont
  {Nozières}}]{PrejbeanuJPCM2007}%
  \BibitemOpen
  \bibfield  {author} {\bibinfo {author} {\bibfnamefont {I.~L.}\ \bibnamefont
  {Prejbeanu}}, \bibinfo {author} {\bibfnamefont {M.}~\bibnamefont {Kerekes}},
  \bibinfo {author} {\bibfnamefont {R.~C.}\ \bibnamefont {Sousa}}, \bibinfo
  {author} {\bibfnamefont {H.}~\bibnamefont {Sibuet}}, \bibinfo {author}
  {\bibfnamefont {O.}~\bibnamefont {Redon}}, \bibinfo {author} {\bibfnamefont
  {B.}~\bibnamefont {Dieny}}, \ and\ \bibinfo {author} {\bibfnamefont {J.~P.}\
  \bibnamefont {Nozières}},\ }\href
  {http://stacks.iop.org/0953-8984/19/i=16/a=165218} {\bibfield  {journal}
  {\bibinfo  {journal} {J. Phys.: Condens. Matter}\ }\textbf {\bibinfo {volume}
  {19}},\ \bibinfo {pages} {165218} (\bibinfo {year} {2007})}\BibitemShut
  {NoStop}%
\bibitem [{\citenamefont {Scholz}\ \emph {et~al.}(2003)\citenamefont {Scholz},
  \citenamefont {Fidler}, \citenamefont {Schrefl}, \citenamefont {Suess},
  \citenamefont {Dittrich}, \citenamefont {Forster},\ and\ \citenamefont
  {Tsiantos}}]{ScholzMAGPAR2003}%
  \BibitemOpen
  \bibfield  {author} {\bibinfo {author} {\bibfnamefont {W.}~\bibnamefont
  {Scholz}}, \bibinfo {author} {\bibfnamefont {J.}~\bibnamefont {Fidler}},
  \bibinfo {author} {\bibfnamefont {T.}~\bibnamefont {Schrefl}}, \bibinfo
  {author} {\bibfnamefont {D.}~\bibnamefont {Suess}}, \bibinfo {author}
  {\bibfnamefont {R.}~\bibnamefont {Dittrich}}, \bibinfo {author}
  {\bibfnamefont {H.}~\bibnamefont {Forster}}, \ and\ \bibinfo {author}
  {\bibfnamefont {V.}~\bibnamefont {Tsiantos}},\ }\href {\doibase
  http://dx.doi.org/10.1016/S0927-0256(03)00119-8} {\bibfield  {journal}
  {\bibinfo  {journal} {Comp. Mater. Sci.}\ }\textbf {\bibinfo {volume} {28}},\
  \bibinfo {pages} {366 } (\bibinfo {year} {2003})},\ \bibinfo {note}
  {proceedings of the Symposium on Software Development for Process and
  Materials Design}\BibitemShut {NoStop}%
\bibitem [{\citenamefont {Fischbacher}\ \emph {et~al.}(2007)\citenamefont
  {Fischbacher}, \citenamefont {Franchin}, \citenamefont {Bordignon},\ and\
  \citenamefont {Fangohr}}]{FischbacherNMAG2007}%
  \BibitemOpen
  \bibfield  {author} {\bibinfo {author} {\bibfnamefont {T.}~\bibnamefont
  {Fischbacher}}, \bibinfo {author} {\bibfnamefont {M.}~\bibnamefont
  {Franchin}}, \bibinfo {author} {\bibfnamefont {G.}~\bibnamefont {Bordignon}},
  \ and\ \bibinfo {author} {\bibfnamefont {H.}~\bibnamefont {Fangohr}},\ }\href
  {\doibase 10.1109/TMAG.2007.893843} {\bibfield  {journal} {\bibinfo
  {journal} {Magnetics, IEEE Transactions on}\ }\textbf {\bibinfo {volume}
  {43}},\ \bibinfo {pages} {2896} (\bibinfo {year} {2007})}\BibitemShut
  {NoStop}%
\bibitem [{\citenamefont {Chang}\ \emph {et~al.}(2011)\citenamefont {Chang},
  \citenamefont {Li}, \citenamefont {Lubarda}, \citenamefont {Livshitz},\ and\
  \citenamefont {Lomakin}}]{ChangFASTMAG2011}%
  \BibitemOpen
  \bibfield  {author} {\bibinfo {author} {\bibfnamefont {R.}~\bibnamefont
  {Chang}}, \bibinfo {author} {\bibfnamefont {S.}~\bibnamefont {Li}}, \bibinfo
  {author} {\bibfnamefont {M.~V.}\ \bibnamefont {Lubarda}}, \bibinfo {author}
  {\bibfnamefont {B.}~\bibnamefont {Livshitz}}, \ and\ \bibinfo {author}
  {\bibfnamefont {V.}~\bibnamefont {Lomakin}},\ }\href {\doibase
  10.1063/1.3563081} {\bibfield  {journal} {\bibinfo  {journal} {J. Appl.
  Phys.}\ }\textbf {\bibinfo {volume} {109}},\ \bibinfo {eid} {07D358}
  (\bibinfo {year} {2011}),\ 10.1063/1.3563081}\BibitemShut {NoStop}%
\bibitem [{\citenamefont {Evans}\ \emph {et~al.}(2014)\citenamefont {Evans},
  \citenamefont {Fan}, \citenamefont {Chureemart}, \citenamefont {Ostler},
  \citenamefont {Ellis},\ and\ \citenamefont {Chantrell}}]{EvansVMPR2013}%
  \BibitemOpen
  \bibfield  {author} {\bibinfo {author} {\bibfnamefont {R.~F.~L.}\
  \bibnamefont {Evans}}, \bibinfo {author} {\bibfnamefont {W.~J.}\ \bibnamefont
  {Fan}}, \bibinfo {author} {\bibfnamefont {P.}~\bibnamefont {Chureemart}},
  \bibinfo {author} {\bibfnamefont {T.~A.}\ \bibnamefont {Ostler}}, \bibinfo
  {author} {\bibfnamefont {M.~O.~A.}\ \bibnamefont {Ellis}}, \ and\ \bibinfo
  {author} {\bibfnamefont {R.~W.}\ \bibnamefont {Chantrell}},\ }\href@noop {}
  {\bibfield  {journal} {\bibinfo  {journal} {J. Phys.: Condens. Matter}\
  }\textbf {\bibinfo {volume} {26}},\ \bibinfo {pages} {103202} (\bibinfo
  {year} {2014})}\BibitemShut {NoStop}%
\bibitem [{\citenamefont {Garanin}(1997)}]{GaraninPRB1997}%
  \BibitemOpen
  \bibfield  {author} {\bibinfo {author} {\bibfnamefont {D.~A.}\ \bibnamefont
  {Garanin}},\ }\href {\doibase 10.1103/PhysRevB.55.3050} {\bibfield  {journal}
  {\bibinfo  {journal} {Phys. Rev. B}\ }\textbf {\bibinfo {volume} {55}},\
  \bibinfo {pages} {3050} (\bibinfo {year} {1997})}\BibitemShut {NoStop}%
\bibitem [{\citenamefont {Evans}\ \emph {et~al.}(2012)\citenamefont {Evans},
  \citenamefont {Hinzke}, \citenamefont {Atxitia}, \citenamefont {Nowak},
  \citenamefont {Chantrell},\ and\ \citenamefont
  {Chubykalo-Fesenko}}]{EvansPRB2012}%
  \BibitemOpen
  \bibfield  {author} {\bibinfo {author} {\bibfnamefont {R.~F.~L.}\
  \bibnamefont {Evans}}, \bibinfo {author} {\bibfnamefont {D.}~\bibnamefont
  {Hinzke}}, \bibinfo {author} {\bibfnamefont {U.}~\bibnamefont {Atxitia}},
  \bibinfo {author} {\bibfnamefont {U.}~\bibnamefont {Nowak}}, \bibinfo
  {author} {\bibfnamefont {R.~W.}\ \bibnamefont {Chantrell}}, \ and\ \bibinfo
  {author} {\bibfnamefont {O.}~\bibnamefont {Chubykalo-Fesenko}},\ }\href
  {\doibase 10.1103/PhysRevB.85.014433} {\bibfield  {journal} {\bibinfo
  {journal} {Phys. Rev. B}\ }\textbf {\bibinfo {volume} {85}},\ \bibinfo
  {pages} {014433} (\bibinfo {year} {2012})}\BibitemShut {NoStop}%
\bibitem [{\citenamefont {Kazantseva}\ \emph {et~al.}(2008)\citenamefont
  {Kazantseva}, \citenamefont {Hinzke}, \citenamefont {Nowak}, \citenamefont
  {Chantrell}, \citenamefont {Atxitia},\ and\ \citenamefont
  {Chubykalo-Fesenko}}]{KazantsevaPRB2008}%
  \BibitemOpen
  \bibfield  {author} {\bibinfo {author} {\bibfnamefont {N.}~\bibnamefont
  {Kazantseva}}, \bibinfo {author} {\bibfnamefont {D.}~\bibnamefont {Hinzke}},
  \bibinfo {author} {\bibfnamefont {U.}~\bibnamefont {Nowak}}, \bibinfo
  {author} {\bibfnamefont {R.~W.}\ \bibnamefont {Chantrell}}, \bibinfo {author}
  {\bibfnamefont {U.}~\bibnamefont {Atxitia}}, \ and\ \bibinfo {author}
  {\bibfnamefont {O.}~\bibnamefont {Chubykalo-Fesenko}},\ }\href {\doibase
  10.1103/PhysRevB.77.184428} {\bibfield  {journal} {\bibinfo  {journal} {Phys.
  Rev. B}\ }\textbf {\bibinfo {volume} {77}},\ \bibinfo {pages} {184428}
  (\bibinfo {year} {2008})}\BibitemShut {NoStop}%
\bibitem [{\citenamefont {Atxitia}\ \emph {et~al.}(2010)\citenamefont
  {Atxitia}, \citenamefont {Hinzke}, \citenamefont {Chubykalo-Fesenko},
  \citenamefont {Nowak}, \citenamefont {Kachkachi}, \citenamefont {Mryasov},
  \citenamefont {Evans},\ and\ \citenamefont {Chantrell}}]{AtxitiaEXC2010}%
  \BibitemOpen
  \bibfield  {author} {\bibinfo {author} {\bibfnamefont {U.}~\bibnamefont
  {Atxitia}}, \bibinfo {author} {\bibfnamefont {D.}~\bibnamefont {Hinzke}},
  \bibinfo {author} {\bibfnamefont {O.}~\bibnamefont {Chubykalo-Fesenko}},
  \bibinfo {author} {\bibfnamefont {U.}~\bibnamefont {Nowak}}, \bibinfo
  {author} {\bibfnamefont {H.}~\bibnamefont {Kachkachi}}, \bibinfo {author}
  {\bibfnamefont {O.~N.}\ \bibnamefont {Mryasov}}, \bibinfo {author}
  {\bibfnamefont {R.~F.}\ \bibnamefont {Evans}}, \ and\ \bibinfo {author}
  {\bibfnamefont {R.~W.}\ \bibnamefont {Chantrell}},\ }\href {\doibase
  10.1103/PhysRevB.82.134440} {\bibfield  {journal} {\bibinfo  {journal} {Phys.
  Rev. B}\ }\textbf {\bibinfo {volume} {82}},\ \bibinfo {pages} {134440}
  (\bibinfo {year} {2010})}\BibitemShut {NoStop}%
\bibitem [{\citenamefont {Asselin}\ \emph {et~al.}(2010)\citenamefont
  {Asselin}, \citenamefont {Evans}, \citenamefont {Barker}, \citenamefont
  {Chantrell}, \citenamefont {Yanes}, \citenamefont {Chubykalo-Fesenko},
  \citenamefont {Hinzke},\ and\ \citenamefont {Nowak}}]{AsselinCMC2010}%
  \BibitemOpen
  \bibfield  {author} {\bibinfo {author} {\bibfnamefont {P.}~\bibnamefont
  {Asselin}}, \bibinfo {author} {\bibfnamefont {R.~F.~L.}\ \bibnamefont
  {Evans}}, \bibinfo {author} {\bibfnamefont {J.}~\bibnamefont {Barker}},
  \bibinfo {author} {\bibfnamefont {R.~W.}\ \bibnamefont {Chantrell}}, \bibinfo
  {author} {\bibfnamefont {R.}~\bibnamefont {Yanes}}, \bibinfo {author}
  {\bibfnamefont {O.}~\bibnamefont {Chubykalo-Fesenko}}, \bibinfo {author}
  {\bibfnamefont {D.}~\bibnamefont {Hinzke}}, \ and\ \bibinfo {author}
  {\bibfnamefont {U.}~\bibnamefont {Nowak}},\ }\href {\doibase
  10.1103/PhysRevB.82.054415} {\bibfield  {journal} {\bibinfo  {journal} {Phys.
  Rev. B}\ }\textbf {\bibinfo {volume} {82}},\ \bibinfo {pages} {054415}
  (\bibinfo {year} {2010})}\BibitemShut {NoStop}%
\bibitem [{\citenamefont {Mryasov}\ \emph {et~al.}(2005)\citenamefont
  {Mryasov}, \citenamefont {Nowak}, \citenamefont {Guslienko},\ and\
  \citenamefont {Chantrell}}]{MyrasovFePtEPL2005}%
  \BibitemOpen
  \bibfield  {author} {\bibinfo {author} {\bibfnamefont {O.~N.}\ \bibnamefont
  {Mryasov}}, \bibinfo {author} {\bibfnamefont {U.}~\bibnamefont {Nowak}},
  \bibinfo {author} {\bibfnamefont {K.~Y.}\ \bibnamefont {Guslienko}}, \ and\
  \bibinfo {author} {\bibfnamefont {R.~W.}\ \bibnamefont {Chantrell}},\ }\href
  {http://stacks.iop.org/0295-5075/69/i=5/a=805} {\bibfield  {journal}
  {\bibinfo  {journal} {Euro. Phys. Lett.}\ }\textbf {\bibinfo {volume} {69}},\
  \bibinfo {pages} {805} (\bibinfo {year} {2005})}\BibitemShut {NoStop}%
\bibitem [{\citenamefont {Kuz'min}(2005)}]{KuzminPRL2005}%
  \BibitemOpen
  \bibfield  {author} {\bibinfo {author} {\bibfnamefont {M.~D.}\ \bibnamefont
  {Kuz'min}},\ }\href {\doibase 10.1103/PhysRevLett.94.107204} {\bibfield
  {journal} {\bibinfo  {journal} {Phys. Rev. Lett.}\ }\textbf {\bibinfo
  {volume} {94}},\ \bibinfo {pages} {107204} (\bibinfo {year}
  {2005})}\BibitemShut {NoStop}%
\bibitem [{\citenamefont {Sandvik}\ and\ \citenamefont
  {Kurkij\"arvi}(1991)}]{SandvikPRB1991}%
  \BibitemOpen
  \bibfield  {author} {\bibinfo {author} {\bibfnamefont {A.~W.}\ \bibnamefont
  {Sandvik}}\ and\ \bibinfo {author} {\bibfnamefont {J.}~\bibnamefont
  {Kurkij\"arvi}},\ }\href {\doibase 10.1103/PhysRevB.43.5950} {\bibfield
  {journal} {\bibinfo  {journal} {Phys. Rev. B}\ }\textbf {\bibinfo {volume}
  {43}},\ \bibinfo {pages} {5950} (\bibinfo {year} {1991})}\BibitemShut
  {NoStop}%
\bibitem [{\citenamefont {Hinzke}\ and\ \citenamefont
  {Nowak}(2011)}]{HinzkePRL2011}%
  \BibitemOpen
  \bibfield  {author} {\bibinfo {author} {\bibfnamefont {D.}~\bibnamefont
  {Hinzke}}\ and\ \bibinfo {author} {\bibfnamefont {U.}~\bibnamefont {Nowak}},\
  }\href {\doibase 10.1103/PhysRevLett.107.027205} {\bibfield  {journal}
  {\bibinfo  {journal} {Phys. Rev. Lett.}\ }\textbf {\bibinfo {volume} {107}},\
  \bibinfo {pages} {027205} (\bibinfo {year} {2011})}\BibitemShut {NoStop}%
\bibitem [{\citenamefont {Bloch}(1930)}]{Bloch1930}%
  \BibitemOpen
  \bibfield  {author} {\bibinfo {author} {\bibfnamefont {F.}~\bibnamefont
  {Bloch}},\ }\href@noop {} {\bibfield  {journal} {\bibinfo  {journal} {Z.
  Physik}\ }\textbf {\bibinfo {volume} {61}},\ \bibinfo {pages} {206} (\bibinfo
  {year} {1930})}\BibitemShut {NoStop}%
\bibitem [{\citenamefont {Bastardis}\ \emph {et~al.}(2012)\citenamefont
  {Bastardis}, \citenamefont {Atxitia}, \citenamefont {Chubykalo-Fesenko},\
  and\ \citenamefont {Kachkachi}}]{BastardisPRB2012}%
  \BibitemOpen
  \bibfield  {author} {\bibinfo {author} {\bibfnamefont {R.}~\bibnamefont
  {Bastardis}}, \bibinfo {author} {\bibfnamefont {U.}~\bibnamefont {Atxitia}},
  \bibinfo {author} {\bibfnamefont {O.}~\bibnamefont {Chubykalo-Fesenko}}, \
  and\ \bibinfo {author} {\bibfnamefont {H.}~\bibnamefont {Kachkachi}},\ }\href
  {\doibase 10.1103/PhysRevB.86.094415} {\bibfield  {journal} {\bibinfo
  {journal} {Phys. Rev. B}\ }\textbf {\bibinfo {volume} {86}},\ \bibinfo
  {pages} {094415} (\bibinfo {year} {2012})}\BibitemShut {NoStop}%
\bibitem [{\citenamefont {Majlis}(2007)}]{QuantumTheoryMagnetism2007}%
  \BibitemOpen
  \bibfield  {author} {\bibinfo {author} {\bibfnamefont {N.}~\bibnamefont
  {Majlis}},\ }\href@noop {} {\emph {\bibinfo {title} {The quantum theory of
  magnetism}}}\ (\bibinfo  {publisher} {McGill University Press, Canada},\
  \bibinfo {year} {2007})\BibitemShut {NoStop}%
\bibitem [{\citenamefont {Garanin}(1996)}]{GaraninPRB1996}%
  \BibitemOpen
  \bibfield  {author} {\bibinfo {author} {\bibfnamefont {D.~A.}\ \bibnamefont
  {Garanin}},\ }\href {\doibase 10.1103/PhysRevB.53.11593} {\bibfield
  {journal} {\bibinfo  {journal} {Phys. Rev. B}\ }\textbf {\bibinfo {volume}
  {53}},\ \bibinfo {pages} {11593} (\bibinfo {year} {1996})}\BibitemShut
  {NoStop}%
\bibitem [{\citenamefont {Stanley}(1968)}]{StanleyPR1968}%
  \BibitemOpen
  \bibfield  {author} {\bibinfo {author} {\bibfnamefont {H.~E.}\ \bibnamefont
  {Stanley}},\ }\href {\doibase 10.1103/PhysRev.176.718} {\bibfield  {journal}
  {\bibinfo  {journal} {Phys. Rev.}\ }\textbf {\bibinfo {volume} {176}},\
  \bibinfo {pages} {718} (\bibinfo {year} {1968})}\BibitemShut {NoStop}%
\bibitem [{\citenamefont {Chikazumi}\ and\ \citenamefont
  {Graham}(1997)}]{chikazumi1997}%
  \BibitemOpen
  \bibfield  {author} {\bibinfo {author} {\bibfnamefont {S.}~\bibnamefont
  {Chikazumi}}\ and\ \bibinfo {author} {\bibfnamefont {C.}~\bibnamefont
  {Graham}},\ }\href {http://books.google.co.uk/books?id=AZVfuxXF2GsC} {\emph
  {\bibinfo {title} {Physics of Ferromagnetism}}},\ International Series of
  Monographs on Physics\ (\bibinfo  {publisher} {Clarendon Press},\ \bibinfo
  {year} {1997})\BibitemShut {NoStop}%
\bibitem [{\citenamefont {Pajda}\ \emph {et~al.}(2001)\citenamefont {Pajda},
  \citenamefont {Kudrnovsk\'y}, \citenamefont {Turek}, \citenamefont {Drchal},\
  and\ \citenamefont {Bruno}}]{PajdaBCCFe2001}%
  \BibitemOpen
  \bibfield  {author} {\bibinfo {author} {\bibfnamefont {M.}~\bibnamefont
  {Pajda}}, \bibinfo {author} {\bibfnamefont {J.}~\bibnamefont {Kudrnovsk\'y}},
  \bibinfo {author} {\bibfnamefont {I.}~\bibnamefont {Turek}}, \bibinfo
  {author} {\bibfnamefont {V.}~\bibnamefont {Drchal}}, \ and\ \bibinfo {author}
  {\bibfnamefont {P.}~\bibnamefont {Bruno}},\ }\href {\doibase
  10.1103/PhysRevB.64.174402} {\bibfield  {journal} {\bibinfo  {journal} {Phys.
  Rev. B}\ }\textbf {\bibinfo {volume} {64}},\ \bibinfo {pages} {174402}
  (\bibinfo {year} {2001})}\BibitemShut {NoStop}%
\bibitem [{\citenamefont {\textsc{vampire}~software package}()}]{vampire-url}%
  \BibitemOpen
  \bibfield  {author} {\bibinfo {author} {\bibnamefont
  {\textsc{vampire}~software package}},\ }\href@noop {} {}\bibinfo {note}
  {Version 3. Available from http://vampire.york.ac.uk/}\BibitemShut {NoStop}%
\bibitem [{\citenamefont {Hinzke}\ and\ \citenamefont
  {Nowak}(1999)}]{HinzkeMC1999}%
  \BibitemOpen
  \bibfield  {author} {\bibinfo {author} {\bibfnamefont {D.}~\bibnamefont
  {Hinzke}}\ and\ \bibinfo {author} {\bibfnamefont {U.}~\bibnamefont {Nowak}},\
  }\href {\doibase http://dx.doi.org/10.1016/S0010-4655(99)00348-3} {\bibfield
  {journal} {\bibinfo  {journal} {Comput. Phys. Commun.}\ }\textbf {\bibinfo
  {volume} {121-122}},\ \bibinfo {pages} {334} (\bibinfo {year}
  {1999})}\BibitemShut {NoStop}%
\bibitem [{\citenamefont {Anderson}\ \emph {et~al.}(1971)\citenamefont
  {Anderson}, \citenamefont {Arajs}, \citenamefont {Stelmach}, \citenamefont
  {Tehan},\ and\ \citenamefont {Yao}}]{Anderson1971}%
  \BibitemOpen
  \bibfield  {author} {\bibinfo {author} {\bibfnamefont {E.}~\bibnamefont
  {Anderson}}, \bibinfo {author} {\bibfnamefont {S.}~\bibnamefont {Arajs}},
  \bibinfo {author} {\bibfnamefont {A.}~\bibnamefont {Stelmach}}, \bibinfo
  {author} {\bibfnamefont {B.}~\bibnamefont {Tehan}}, \ and\ \bibinfo {author}
  {\bibfnamefont {Y.}~\bibnamefont {Yao}},\ }\href {\doibase
  http://dx.doi.org/10.1016/0375-9601(71)90407-5} {\bibfield  {journal}
  {\bibinfo  {journal} {Phys. Lett. A}\ }\textbf {\bibinfo {volume} {36}},\
  \bibinfo {pages} {173 } (\bibinfo {year} {1971})}\BibitemShut {NoStop}%
\bibitem [{\citenamefont {Anisimov}\ \emph {et~al.}(1974)\citenamefont
  {Anisimov}, \citenamefont {Kapeliovich},\ and\ \citenamefont
  {Perelman}}]{AnisimovTTM1974}%
  \BibitemOpen
  \bibfield  {author} {\bibinfo {author} {\bibfnamefont {S.}~\bibnamefont
  {Anisimov}}, \bibinfo {author} {\bibfnamefont {B.}~\bibnamefont
  {Kapeliovich}}, \ and\ \bibinfo {author} {\bibfnamefont {T.}~\bibnamefont
  {Perelman}},\ }\href@noop {} {\bibfield  {journal} {\bibinfo  {journal} {Sov.
  Phys. JETP}\ }\textbf {\bibinfo {volume} {39}},\ \bibinfo {pages} {375–377}
  (\bibinfo {year} {1974})}\BibitemShut {NoStop}%
\bibitem [{\citenamefont {Lin}\ \emph {et~al.}(2008)\citenamefont {Lin},
  \citenamefont {Zhigilei},\ and\ \citenamefont {Celli}}]{LinPRB2008}%
  \BibitemOpen
  \bibfield  {author} {\bibinfo {author} {\bibfnamefont {Z.}~\bibnamefont
  {Lin}}, \bibinfo {author} {\bibfnamefont {L.~V.}\ \bibnamefont {Zhigilei}}, \
  and\ \bibinfo {author} {\bibfnamefont {V.}~\bibnamefont {Celli}},\ }\href
  {\doibase 10.1103/PhysRevB.77.075133} {\bibfield  {journal} {\bibinfo
  {journal} {Phys. Rev. B}\ }\textbf {\bibinfo {volume} {77}},\ \bibinfo
  {pages} {075133} (\bibinfo {year} {2008})}\BibitemShut {NoStop}%
\end{thebibliography}%

\end{document}